# #ILookLikeAnEngineer: Using Social Media Based Hashtag Activism Campaigns as a Lens to Better Understand Engineering Diversity Issues

**Dr. Aqdas Malik, George Mason University**
**Dr. Aditya Johri, George Mason University**

Aditya Johri is Associate Professor in the department of Information Sciences & Technology. Dr. Johri studies the use of information and communication technologies (ICT) for learning and knowledge sharing, with a focus on cognition in informal environments. He also examine the role of ICT in supporting distributed work among globally dispersed workers and in furthering social development in emerging economies. He received the U.S. National Science Foundation's Early Career Award in 2009. He is co-editor of the Cambridge Handbook of Engineering Education Research (CHEER) published by Cambridge University Press, New York, NY. Dr. Johri earned his Ph.D. in Learning Sciences and Technology Design at Stanford University and a B.Eng. in Mechanical Engineering at Delhi College of Engineering.

**Mr. RAJAT HANDA**
**Mr. Habib Karbasian, George Mason University**

PhD student in IT

**Dr. Hemant Purohit, George Mason University**

# #ILookLikeAnEngineer: Using Social Media Based Hashtag Activism Campaigns as a Lens to Better Understand Engineering Diversity Issues

**Abstract**

Each year, significant investment of time and resources is made to improve diversity within engineering across a range of federal and state agencies, private/not-for-profit organizations, and foundations. In spite of decades of investments, efforts have not yielded desired returns - participation by minorities continues to lag at a time when STEM workforce requirements are increasing. In recent years a new stream of data has emerged - online social networks, including Twitter, Facebook, and Instagram - that act as a key sensor of social behavior and attitudes of the public. Almost 87% of the American population now participates in some form of social media activity. Consequently, social networking sites have become powerful indicators of social action and social media data has shown significant promise for studying many issues including public health communication, political campaign, humanitarian crisis, and, activism. We argue that social media data can likewise be leveraged to better understand and improve engineering diversity. As a case study to illustrate the viability of the approach, we present findings from a campaign, #ILookLikeAnEngineer (using Twitter data – 19,354 original tweets and 29,529 retweets), aimed at increasing gender diversity in the engineering workplace. The campaign provided a continuous momentum to the overall effort to increase diversity and novel ways of connecting with relevant audience. Our analysis demonstrates that diversity initiatives related to STEM attract voices from various entities including individuals, large corporations, media outlets, and community interest groups.

## Introduction

The term "STEM education" refers to teaching and learning in the fields of science, technology, engineering, and mathematics. According to Sanders (2008), in the 1990s, the National Science Foundation (NSF) "SMET" was the shorthand for "science, mathematics, engineering, and technology" and an NSF program officer complained that "SMET" sounded too much like "smut," that resulted in the new acronym "STEM" [1]. Although it took some time for STEM to catch on, and even as recently as 2003 few people know what it meant, in the last decade the use of the term has exploded [2]. STEM is now used to refer to educational activities across all grade levels (from pre-school to post-doctorate) and to both formal (e.g., classrooms) and informal (e.g., after school programs) experiences. According to data available publicly, annual federal appropriations for STEM education are now in above $4 billion. There is an active interest in STEM education at that level given the importance placed on advancing science and engineering to maintain U.S. competitive edge. STEM education is now a topic used frequently in policy debates, including issues of education, workforce development, national security and immigration. According to the Congressional Research Service STEM primer [3], there are between 105 and 252 STEM education programs or activities at 13 to 15 federal agencies; the key agencies involved in the effort are Department of Education, National Science Foundation, and Health and Human Services.

Given that the interests are broad and federal efforts are spread across multiple agencies, there is a concern with perceived duplication of effort and a lack of coordination in the federal effort. Therefore, efforts to improve accountability and coordination have gained prominence in recent years. The data currently available about STEM education paints a "complicated" picture [3]. According to many indicators [3], overall graduate enrollments in science and engineering (S&E) grew 35% over the last decade; S&E enrollments for Hispanic/Latino, American Indian/Alaska Native, and African American students (all of whom are generally underrepresented in S&E) grew by 65%, 55%, and 50%, respectively. Yet, concerns remain about persistent academic achievement gaps between various demographic groups, STEM teacher quality, the rankings of U.S. students on international STEM assessments, foreign student enrollments and increased education attainment in other countries, and the ability of the U.S. STEM education system to meet domestic demand for STEM labor.

To better understanding STEM efforts, different kinds of data sources are used, such as National Assessment of Educational Progress (NAEP), Trends in International Mathematics and Science Study (TIMSS), Program for International Student Assessment (PISA), Science and Engineering Indicators (S&E), and NCSES, among others. GAO and other federal offices also periodically release information that assists in understanding STEM issues. The use of surveys is the preferred method to produce data about different programs and initiatives both due to the convenience of data collection and because the data collection and analysis is consistent with statistical measures that have been perfected over decades. The data and analysis generated here is reliable and useful. But, as Feuer (2013) has argued, in spite of this data and the effort to better understand STEM outcomes "the connections between diagnosis and prescription were, and remain, logically and empirically challenging (p. 3)." **Traditional data are limited in what they tell us about STEM education.** In particular, they do not capture the pulse of the people nor provide insight into the major entities that can assist STEM initiatives. They are, like most STEM initiatives, top-down and very strongly tied to activities in formal institutions. Consequently, there has been a renewed interest in recent years to revisit the limitations and quality of existing data and, as Feuer (2013) argues, "the need for more and better data, and priority topics for a sustained research agenda (p. 5)" [4].

These newer data and information are needed especially if we want to pursue and achieve many of the goals outlined by policy researchers that are shown to work. For instance, Malcom-Piqueux & Malcom (2013) argue that greater awareness of careers, role models, after- and out-of-school experiences are critical for equity. They suggest that "students and their families need encouragement and access to information at a much earlier stage than has typically been provided, through exposure to role models who look like them, information about the kinds of jobs done by persons with preparation in engineering, and examples to dispel the idea that engineering is solitary work (pg. 32) [5]." They further contend that efforts to increase the participation of females and underrepresented minorities in engineering education and careers started outside of school, through programs such as Expanding Your Horizons for female students and Mathematics, Engineering, Science Achievement (MESA), Sally Ride Science and AAAS' Spark Club (part of a suite of programs under NSF's Innovative Technology Experiences for Students and Teachers, ITEST), After School Alliance, and others but few parents have awareness of these programs and communities must be engaged and relevant contexts created for students from different populations. All the issues outlined here are easy to address if we have the reach – access to students and their families – and if we are able to provide them with the right information at the right time. Furthermore, the efforts we make will be even more fruitful if we are able to construct channels that are reusable, through which we can easily access the right target population, and which are natural for our intended audience.

In recent years, with the advent of the Internet, new sources of data about public behavior and discourse of public interactions are increasingly becoming available. This increase in the quantity and quality of data has occurred at the same time as the capacity to collect and store data as well the ability to analyze this data have improved significantly. One significant source of data for understanding events and activities that are of concern and of issues that matter to us is data generated from social networking sites or social media. Social media include Internet-based platforms such as Facebook, Instagram, and Twitter, among others, which continue to gain users due to increase in mobile devices based user interaction through different apps that are available as part of the device ecosystem. Almost 87% of the American population now participates in some form of social media activity. Social media provide powerful indicators of social action as they allow individual and small groups to find like-minded individuals and organizations and topics around which they can coalesce. Social media data has been especially useful for researchers studying issues around health communication [6], politics [7], humanitarian crisis [8], social movements such as Arab Spring and Black Lives Matters [9], [10] and activism [11]. For example, Achrekar et al. (2011) has exploited Twitter data for a real-time assessment to predict flu. The model is trained and tested with the data collected by U.S. Center for Disease Control and Prevention, and shows that Twitter data can improve the model prediction accuracy [12]. These efforts suggest that social media is an appropriate channel for better understanding STEM education issues.

STEM education has a key implication for the workforce development and thus, the distribution of STEM workforce can be a reflection of the state of STEM education in the society.

Unfortunately, the recent events and campaigns for stereotyping in the workforce indicate the poor diversity of STEM workers and how the minorities are stereotyped – particularly women. This behavior further undermines the efforts to raise awareness for STEM education. On the other hand, there are several initiatives by established organizations to raise awareness for workforce diversity and thus, implicitly promoting STEM education. In this paper we draw on content analysis and social network analysis methods to conduct an exploratory study of STEM workforce stereotyping and diversity issues using social media activities, and participating audience in a popular campaign. We apply data, techniques, expertise, and interdisciplinary perspectives that have not been applied to the STEM domain by exploring a STEM diversity initiate – #ILookLikeAnEngineer campaign. Using a bottoms-up approach and looking at informal influences on STEM education by means of awareness for workforce diversity, including the media, public, and the businesses, we build a more comprehensive and nuanced picture.

The fundamental contribution of this research is in shedding light on issues that have the potential to impact multiple stakeholders and how they pursue STEM related efforts. For instance, for projects looking for STEM partners in a specific geographical area, our research can provide that information. There is an opportunity to aggregate information about scholarships and competitions as they are often announced through social media. Overall this exploratory research can help build a system that will assist in raising awareness of STEM issues, better understanding of issues, better understanding of sentiment regarding issues, figure out geographic distribution, provide temporal understanding of issues, and help understand communities and networks – who is connected and who can be connected. For instance it will provide an avenue to share real world examples [13]; allow participants to share their STEM related stories [14]; and, engender better debates around STEM issues [15].

Hashtag Activism Campaign: #ILookLikeAnEngineer

In August 2015 a hashtag on Twitter, #ILookLikeAnEngineer, started trending highly and attracted a lot of media coverage and participation by users. The hashtag was born out of one woman's frustration with how people reacted to a recruitment advertisement by her company One Login that featured her. In late July 2015 the company OneLogin posted billboards across public transport in the California Bay Area, especially at the BART train stations, as a recruitment drive (see Fig. 1). One of the billboards depicted a woman engineer Isis Wenger (now Isis Anchalee) and her photo attracted a lot of attention on the Web. Her image led to discussions online about the veracity of the campaign as some people found it unlikely that she was really an engineer.

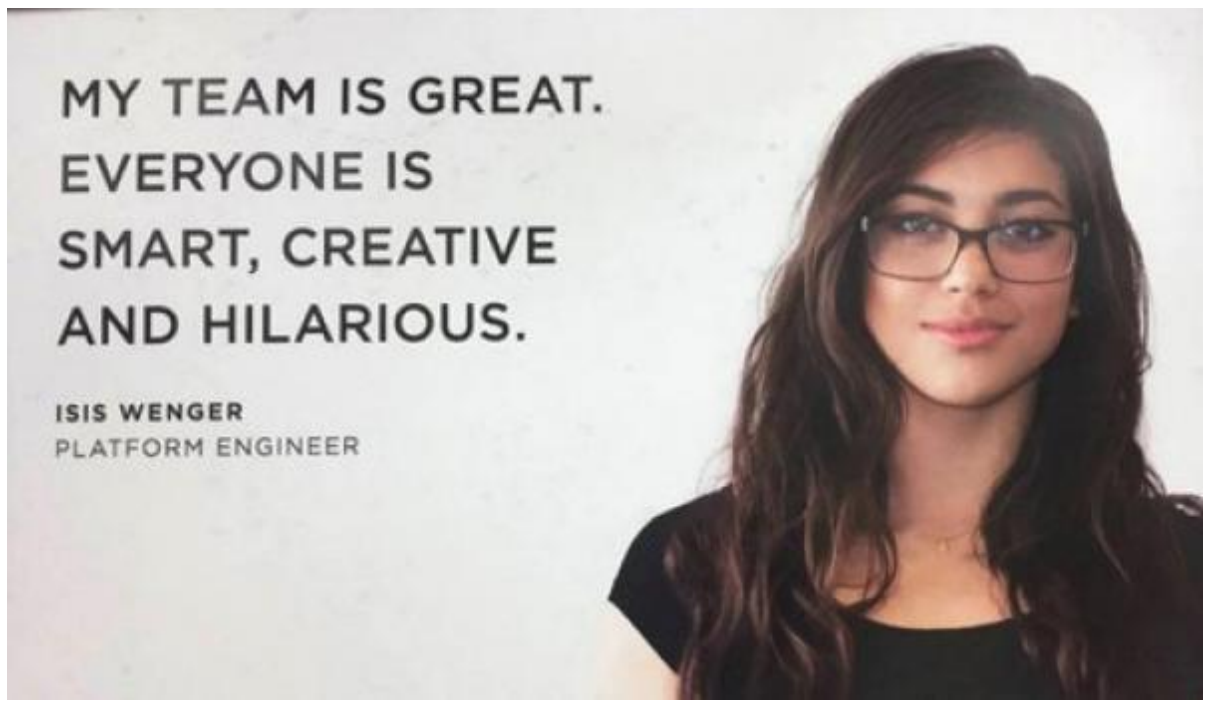

Fig. 1. OneLogin billboard at a BART station

Online comments stated that she was "too attractive" to be a "real engineer," among a lot of other demeaning comments. These online discussions, according to Anchalee, prompted her to write a post on Medium, on August 1, 2015, responding to the stereotypical reaction her image in the ad had generated. She stated in the post that, "At the end of the day, this is just an ad campaign and it is targeted at engineers. This is not intended to be marketed towards any specific gender — segregated thoughts like that continue to perpetuate sexist thought-patterns in this industry." She wanted to show that she was not the only female engineer and also take some of the attention away from her and towards the issue of diversity in engineering. Her Medium post as well the hashtag received large support and the Twitter tag in particular took off as a way for others to share their images and their ideas on the issue. The hashtag soon saw significant media coverage not just in the US but also across the Atlantic, particularly in the UK. The movement not only had an online presence but a large event was organized in September 2015 in the Bay Area supported by a group. The hashtag movement was started as a campaign to raise funds to be able to post billboards across the Bay Area with the hashtag and photos; in effect, to take the online movement into the physical world through a series of billboards.

## Research Study

### *Analytical Foci, Research Questions, and Overview of Approach*

Our exploratory research combines existing research on STEM Education issues, such as broadening participation, with new techniques in the emerging field of computational social science [16] for analysis of data, for visualization of data, and for overall improving how data can be used. Our early-concept has two analytical foci (see Table 1 below for details):

1) *Actors*: Who are the participants in STEM workforce diversity related tweets? For this question, we use a qualitative coding scheme to categorize types of participating actors, and study patterns of their interaction in the network.

2) *Activity*: What activities do they engage in and what is the distribution of these activities? For this question, we use geographical/spatial-temporal analysis to gauge the interest towards STEM workforce diversity efforts and campaigns related messages from specific regions, over time.

Table 1. Overview of Research Study – Analytical Foci, Expected Contribution, and Approach

| **Analytical Foci and Purpose** | **Contribution to Improve Understanding of STEM Issues** |
|---|---|
| ***Actors*** | |
| *Who participates?* Identify primary individuals and organizations participating in STEM issues. | Establish who is participating and who is absent – individuals as well as organizations. Understand the nature of participation and self-representation of participants in order to leverage them for interventions. |
| ***Activities*** | |
| *What do participants do?* Identify major STEM related activities, initiatives, and opportunities. | It is important that any activity or opportunity is appropriately targeted and there is outreach around it. It is also important to know, geographically and in terms of topics, what is going on when and where. |

***Research Data***

For this research we use data from the social network site Twitter - a highly popular microblogging platform. Twitter, in addition to Facebook, has become the de facto tool for direct sharing of information by members of the public and by organizations. Through the use of hashtags, tweeting, and retweeting, Twitter allows users to voice their concerns, connect with others and show solidarity, while also demonstrating their personal and/or organizational identities.

Our overall data corpus though will be larger than just the number of tweets as we will also collect publicly available information referenced in the tweets. For instance, users often refer to external webpages and images which can provide useful information about the topic. There is also some limited but useful information available about what users and the public at large searches for on the Web that can be usefully linked to data from social media. This part is quite important as in addition to the words in the tweets, the links when expanded lead to more important information. In spite of the advances in data science, there are multiple challenges in working with any form of data, including Twitter data, and we are well aware of some of the issues we might face. For instance, the data from Twitter are often noisy, grammatically incorrect, and short in length, which challenges the direct application of standard text analytics methods. To undertake meaningful analysis of such data is not straightforward. For our analysis we will leverage studies in other domains as many relevant techniques have been developed and we will develop some domain specific algorithms if required.

We initiated the data collection for the dataset through Twitter streaming API, by seeding three hashtags for crawling: #ILookLikeAnEngineer, #LookLikeAnEngineer, and #LookLikeEngineer. The time frame for the data ranged from August 3rd, 2015 (the day hashtag was first used) till October 15th, 2015. In order to have a reliable and stabilized retweet count, we recollected the tweets through Twitter search API after a year. The final dataset consists of 19,354 original tweets and 29,529 retweets. Along with tweet text, the following metadata was also collected: retweet count, favorite count, time of tweet; for the user: name, screen name, location, followers, following, likes, and number of tweets.

***Analysis***

Even though our intent in this proposed work is to undertake research which is exploratory and risky, to test the viability of our approach we conducted a preliminary analysis. *Exploratory research questions* that guided the preliminary study included: a) Participation: Who are the most active users in terms of follower count in the ILookLikeAnEngineer (ILLAE) dataset? Who are the most active users in terms of tweet count? Where are the tweets coming from? b) Discourse: What are the most frequent hashtags and their associated themes? c) Interaction: What is a pattern of user interaction and its dynamics? In the following sections we present the detailed analysis of these inquiries.

## Participation

***User analysis:***

To understand the drivers of the campaign, we looked at the aspects related to highly significant tweeters with respect to number of tweets, retweets, and favorites count. In the current dataset, 13,270 unique users contributed through their tweets generating a total of 29,529 retweets and 43,450 favorites. In order to gain a better understanding of the prominent users within the network, we specifically focused on the top 25 users. Statistics of the user profile also provided insights that lead to a better understanding of the drivers of the campaign and their intentions. We further classified each of the respective Twitter profile manually by browsing through the Twitter page.

Analysis of the most active users indicates active involvement of female professionals (mostly engineers) and large engineering companies. Seven women were in the top 25 users with highest number of tweets. Michelle Glauser who was the campaign manager of #ILookLikeAnEngineer, followed by a Female IT engineer from India tweeted the most throughout the campaign (see Table 2). The campaign originator Isis Anchalee as well as women engineers from large corporations such as GE and Cisco also contributed actively to the campaign. From the corporate side, GE was one of the leading entities as they actively tweeted using five different Twitter accounts: GEConnections, GECareers, GeneralElectric, GEHiresHeroes, and GEEuropeCareers. Other engineering companies active in the campaign included Caterpillar and Cisco.

Table 2. Users posting highest number of tweets

| Twitter handle | User type | No. of tweets | No. of Followers |
|---|---|---|---|
| Michelle Glauser | Female | 106 | 3845 |
| Aditi Kapoor | Female | 89 | 317 |
| Richa Sachdevaa | Female | 65 | 182 |
| Isis Anchalee | Female | 56 | 20017 |
| GEConnections | Company | 52 | 17266 |
| Sitesetup | Company | 47 | 1132 |
| Shaunda | Female | 45 | 710 |
| Caterpillar Inc. | Company | 37 | 96244 |
| Hackathon News | Company | 36 | 2039 |
| Savy Nic | Male | 32 | 338 |
| SNC-Lavalin | Company | 31 | 16524 |
| GECareers | Company | 31 | 22738 |
| General Electric | Company | 27 | 426703 |
| GEHiresHeroes | Company | 26 | 2659 |
| We Are Cisco | Company | 24 | 22540 |
| Brian Ballantyne | Male | 24 | 516 |
| GE Europe Careers | Company | 24 | 7571 |
| Zzadet | Individual | 24 | 104 |
| Suzanne Defache | Female | 24 | 390 |
| Nasser Sahlool | Male | 23 | 596 |
| ThoughtWorks | Company | 23 | 161701 |
| Nevin Weston | Female | 23 | 242 |
| CiscoWISE | Company | 21 | 1128 |
| BookBuddi | Company | 20 | 285 |
| Stockmarketnow | Company | 20 | 971 |

Even though organizations participated actively in terms of the number of tweets (see Table 2), their presence was limited in other indices of engagement. In spite of a large number of followers only six organizations make up in the list of top 20 retweets (see Table 3). This finding indicates the limited impact of corporates in engaging users to retweet their content within the campaign. On the other hand, tweets by individuals were retweeted substantially more than other entities. Individuals accounted for 60% of the most retweeted tweets highlighting the influence of individuals in hashtag activism campaigns. Though some of these individuals are well known such as Scott Kelly, Tracy Chou, and Isis Anchalee (campaign founder), yet many of the users are not (e.g. Marcos Caceres and Jolene Hayes).

Further examination of corporations in the list reveals that out of the six corporations identified, five are media outlets. We also analyzed the ratio of verified and non-verified entities to assess their respective level of engagement. The analysis indicates that slightly over half of the accounts were verified (55%), and the rest of them (45%) were non-verified accounts. In general, non-verified users on Twitter have a limited reach in terms of number of followers, yet within this campaign their tweets were retweeted proactively that signifies their importance in propagating the message in such campaigns. One of the reasons behind this phenomenon might be that tweets from non-verified users (who are generally individuals) have an emotional tone in their tweets. Furthermore, it is also plausible that the users who retweeted tweets of non-verified individuals might have associated themselves personally and emotionally with the campaign.

Table 3. Users receiving highest number of retweets

| Twitter handle | User type | Verified/Non-verified | Tweet type* | No. of retweets |
|---|---|---|---|---|
| Dara Oke | Female | NV | T+P | 2450 |
| Marcos Caceres | Male | NV | T+P | 2088 |
| Keyz MIning | Male | NV | T+L | 1999 |
| Isis Anchalee | Female | NV | T+P | 898 |
| TechCrunch | Company | V | T+L | 803 |
| The New York Times | Company | V | T+L | 771 |
| BuzzFeed | Company | V | T+P+L | 715 |
| Tesla | Company | V | T+P | 713 |
| MIT | University | V | T+P | 710 |
| Tracy Chou | Female | V | T+P | 670 |
| Scott Kelly | Male | V | T+P | 660 |
| Mashable | Company | V | T+P+L | 659 |
| BuzzFeed | Company | V | T+P+L | 654 |
| Kaya Thomas | Female | V | T+P | 588 |
| Erica Joy | Female | V | T+P | 586 |
| Pamela Assogba | Female | NV | T+P | 552 |
| SciencePorn | Community | NV | T+P+L | 512 |
| Lisa Smith | Female | NV | T+P | 494 |
| Holly | Female | NV | T+P | 491 |
| Jolene Hayes | Female | NV | T+P | 450 |

*T=text, P=photo, L=Link

Likewise retweets count, the favorites count also shows a comparable result. The list of 20 most favorited tweets indicates that 60% of them were tweeted by individuals (see Table 4). Organization, media, NGO/community, and university hold rest of the 40%. Moreover, tweets of verified and non-verified accounts are favorited almost equally. Overall our findings from the most favorited and retweeted tweets are in agreement that, for the online discourse related to this campaign individuals (ordinary citizens and to some extent notable personalities) dominated over other entities including organizations and media outlets.

Table 4. Users receiving highest number of favorites

| **Twitter handle** | **User type** | **Verified/Non-verified** | **Tweet type** | **No. of favorites** |
|---|---|---|---|---|
| Marcos Caceres | Male | NV | T+P | 3378 |
| Dara Oke | Female | NV | T+P | 2725 |
| Scott Kelly | Male | V | T+P | 2011 |
| Tracy Chou | Female | V | T+P | 1427 |
| Tesla | Company | V | T+P | 1373 |
| Isis Anchalee | Female | NV | T+P | 1264 |
| Erica Joy | Female | V | T+P | 1264 |
| BuzzFeed | Company | V | T+P+L | 964 |
| Jolene Hayes | Female | NV | T+P | 954 |
| Lisa Smith | Female | NV | T+P | 940 |
| BuzzFeed | Company | V | T+P+L | 927 |
| Emily Calandrelli | Female | V | T+P+L | 912 |
| Kaya Thomas | Female | V | T+P | 900 |
| Snapchat | Company | V | T+P | 889 |
| Pamela Assogba | Female | NV | T+P | 888 |
| SciencePorn | Community | NV | T+P+L | 861 |
| Texas Instruments | Company | NV | T+P | 839 |
| Ford Motor Company | Company | V | T+P | 803 |
| MIT | University | V | T+P | 797 |
| Sailor Mercury | Female | NV | T+P | 772 |

*T=text, P=photo, L=Link

In order to get a better sense of participating entities within the campaign, we further examined the engagement level of various entities. We analyzed the 284 most retweeted user profiles out of 13270 unique user profiles (~2.14%). These user profiles were classified into five categories: individuals (e.g. Isis Anchalee, Tracy Chou, and Erica Joy), companies (e.g. Tesla, Google, and Microsoft), news and media outlets (e.g. CNN, BBC, and Independent), non-government organizations/communities (e.g. Code for America, Care.org, and Women's Engineering Society) and universities (e.g. MIT, UC San Francisco, and Victoria). The results from this analysis further reaffirm our previous finding as an overwhelming majority of individuals compose this list (42.6%) followed by media/news outlets (22.5%), and organizations (19.4%).

Further analysis indicates that among the 284 most retweeted users over 46% (N=131) are non-verified. Despite a general assumption that non-verified users are not public figures or well-known organizations, and have far limited number of followers, they were actively retweeted and favorited. This finding also endorses the notion that this campaign has been driven by individuals who identified with the core idea behind the movement. On the other hand, involvement of verified users was equally significant for the success of the campaign. A large number of corporates and news/media outlets participating in this campaign have verified accounts indicating their support in wider propagation of the campaign message.

Geolocation analysis:

Analyzing the tweets location further helped us in gauging the engagement level of the users based on their location. A limited number of tweets (0.77%) in the collected dataset had geolocation metadata. Therefore, in order to have a better understanding of the location of users within our dataset, we captured the location of the users through location information their Twitter profiles. Out of the total of 19,354 original tweets, around 25% of the user profiles did not display the location. Overall 2600 unique locations were geocoded through Google Map API. The granularity of the results goes up to the county, state, or city level depending on how detailed the location is provided by the user. By parsing the JSONs, we were able to find the coordinates of the user locations that are displayed in the world map (see Fig. 2).

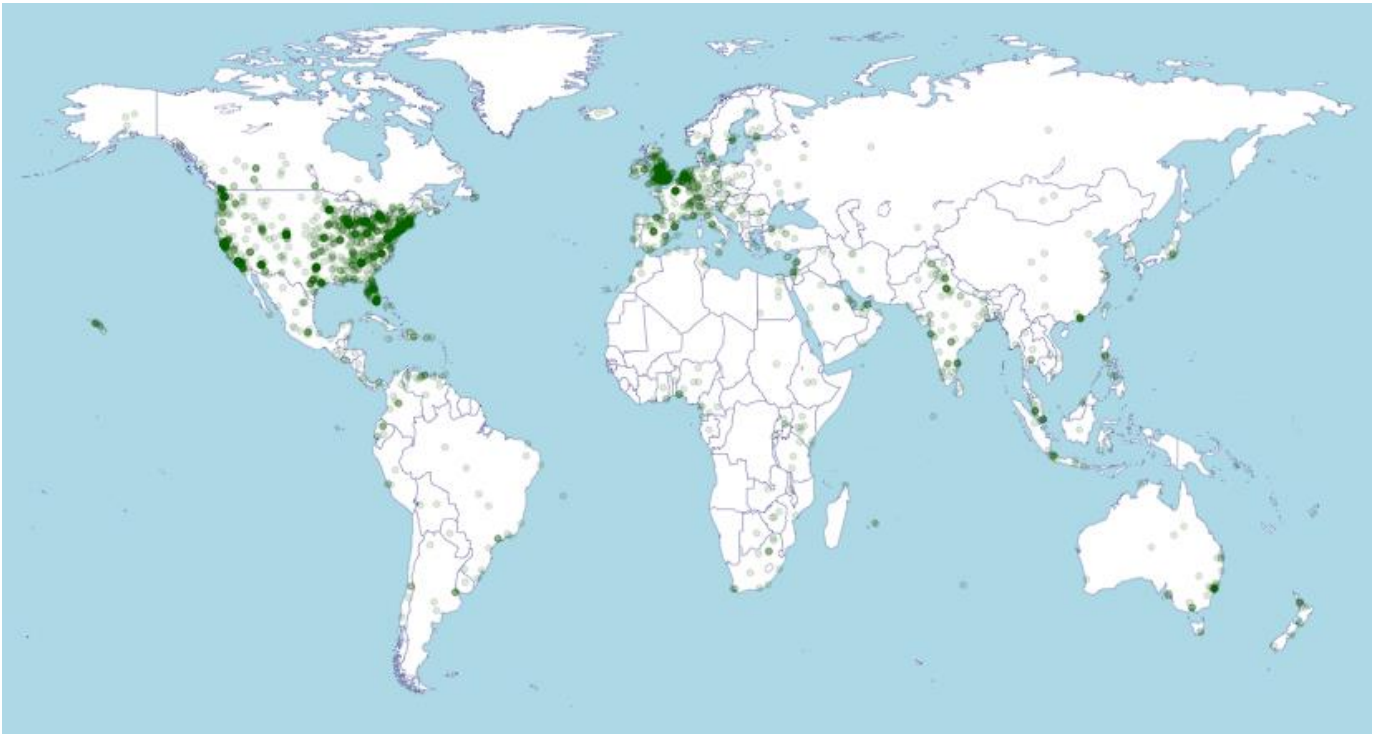

Fig. 2. Tweet activity across the globe

Analyzing the geolocation indicates that users across the globe participated in the campaign (138 countries). An overwhelming majority of the tweets (64%) were from US-based users. Twitter users from Canada, UK, and European countries (predominantly France and Switzerland) with a relatively large engineering and technical presence also actively participated in the campaign. As majority of the tweets were from US-based users, we limited our detailed analysis to state level. The analysis indicate that majority of the tweets were sent from east and west coast, particularly from the states where most of the high-tech companies are based.

## Campaign discourse

### *Words and hashtags analysis*

To understand the discussion topics and themes of interest dueing the campaign, we carried out words and hashtags analysis. For words analysis, we utilized Latent Dirichlet Allocation (LDA) which is an unsupervised machine learning technique [17]. This technique has been often used and considered highly suitable for identifying topics related to tweets. We used log likelihood of the model for each topic. We found out that log likelihood for six topics was the highest and was sufficient to see what all is being discussed related to hashtag. As a complement to the words analysis, we also utilzed the term frequency analysis and visulaized it with the top term used within the tweets (see Fig. 3).

Results from LDA technique revealed five key discussion themes. The main topic of interest as captured by keywords: *community, tech, inspiring, hosted,* and *powerful* refer to the event that was hosted on August 13 to raise funds for the campaign billboards to be erected in San Francisco. A major chunk of the tweets related to the announcement and participation in the event. People were inspired by the event as six women speakers (including Isis Anchalee) expressed their experiences as underrepresented minorities in the tech workforce. A number of tweets were also framed towards increasing the public awareness and highlighting the views, information and attitudes around the campaign. This aspect was expressed through the tweet terms such as: *tech, work, challenge, software,* and *share*.

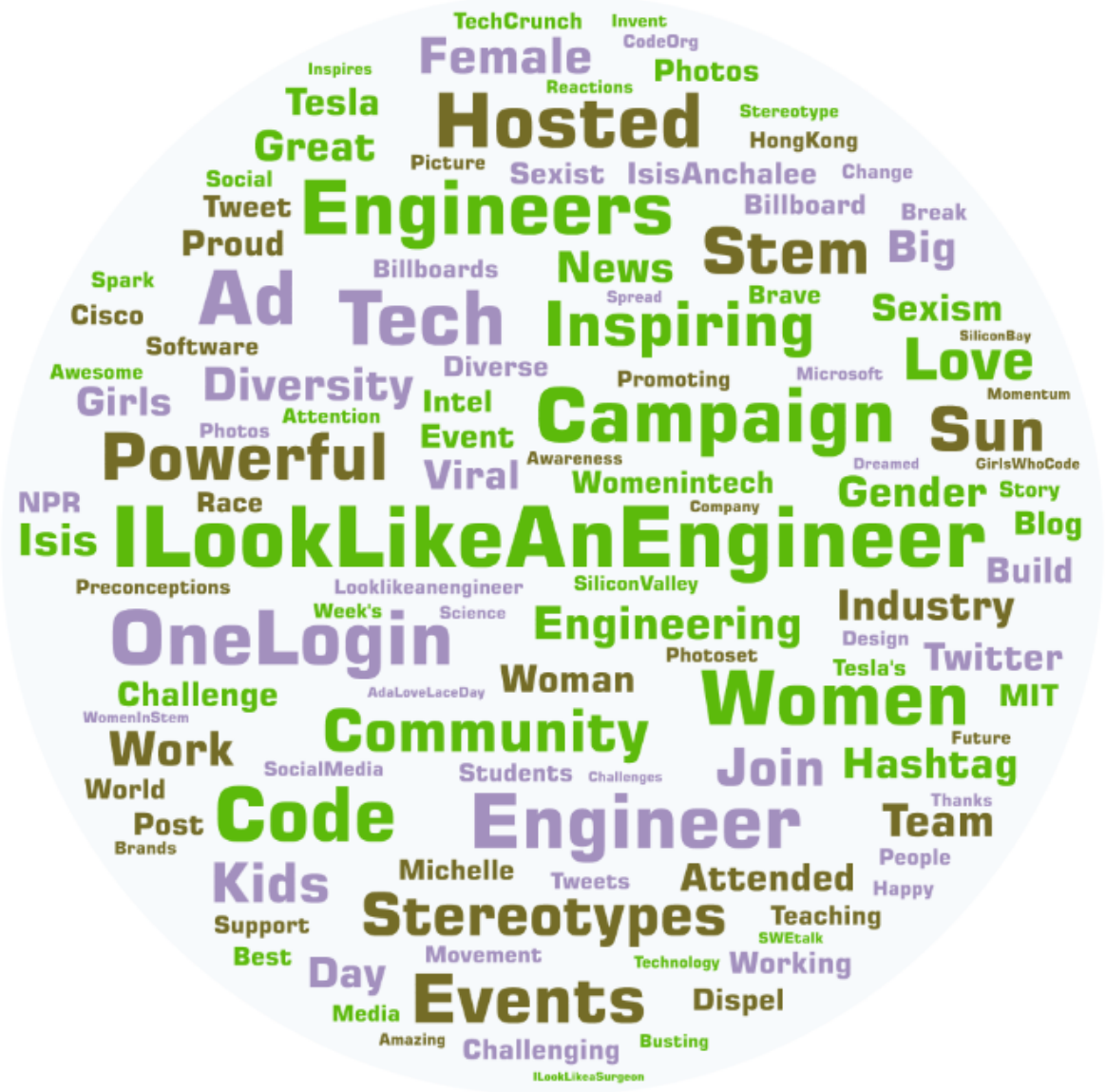

Fig. 3. Keywords used in #ILLAE tweets.

Showcasing the contributions of female engineers and call for diversity in the engineering workforce also appeared as a major theme discussed and followed during the campaign. Using the hashtag #iLookLikeAnEngineer, many users (predominantly women) tweeted and posted pictures of themselves along with a description of their role in the engineering fields. These views were captured through words including *campaign, diversity, gender, challenging,* and *aims*. The hashtag originator, Isis Anchalee also received enormous support and appreciation for initiating the campaign. Women from various fields, particularly STEM encouraged other users to participate and raise public awareness by tweeting the hashtag. These sentiments were presented through terms including *female, womenintech, Twitter, great,* and *women*. Finally, the key aims of the campaign that include challenging stereotypes in engineering domain as well as erecting billboard in the Bay area also appeared as highly debated and followed topic. Tweets around this topic accompanied terms: *stereotypes, STEM, working, billboard,* and *viral*.

The hashtags analysis also reiterate a similar pattern as generated through words analysis. The campaign hashtag #Ilooklikeanenginner was obviously the leading hashtag used within the tweets.

The involvement and role of women in STEM workforce was expressed through hashtags #womenintech, #womeninstem, #women, and #Adalovelaceday. The other leading hashtags used throughout the campaign were #diversity, #race, STEM, tech, and engineering that echo the tweets calling for diversity in the STEM education and workforce.

### Campaign interaction

***Tweet type analysis:***
A detailed analysis of the content modality of tweets indicates a very interesting pattern. Almost half of the original tweets contain text and link, followed by tweets that contain text and a photo. In total around two-third of the original tweets fall in this category, indicating the importance of photo and link in addition to tweet text. Appending link and photo point out that beside text additional information is provided that might not be possible otherwise due to text limitation. Further examination of retweets and favorites showed that tweets containing a photo stand clearly apart from other modalities.

Although only around 40% of the original tweets contain an image, majority of the highly retweeted tweets (85%) are embedded with a photo. This finding signifies the importance of embedding photos in the tweets that can have a significant impact of the retweet-ability. Likewise, all of the tweets in the most favorited list contained a photo designating a significant impact of photos on favorite-ability. This finding further endorses our inference that visual content in tweets are more appealing and perceived more positively than the other form of content or simply textual tweets by the users. Furthermore, this outcome aligns and reaffirms the power of photos in contemporary communication, especially in the context of social media where photos are used not only for social influence and disclosure but for information sharing purposes as well [18]. In the context of this campaign, an overwhelming majority of photos are of female engineers working in their workplaces or having routine daily activities e.g. spending time with family. As these pictures are closer to reality and many people can somewhat relate to them personally or someone they know. Furthermore, these picture also portray the contributions of female engineers in the workforce. Although this finding is confined to the current case, but it might be applicable to other datasets.

***Network analysis:***

To further understand the significant users within the community, we conducted network analysis based on the retweets. Each node represented a user (either retweeted or got retweeted) and the directed edge denoted the relationship between two nodes (users) from a user who retweets to another user. For creating the network graph and measuring various network centralities, we employed Gephi – an open source network visualization tool [19]. The #ILLAE network is composed of 24702 nodes and 27442 edges. Network visualization of the campaign is presented in Fig. 4.

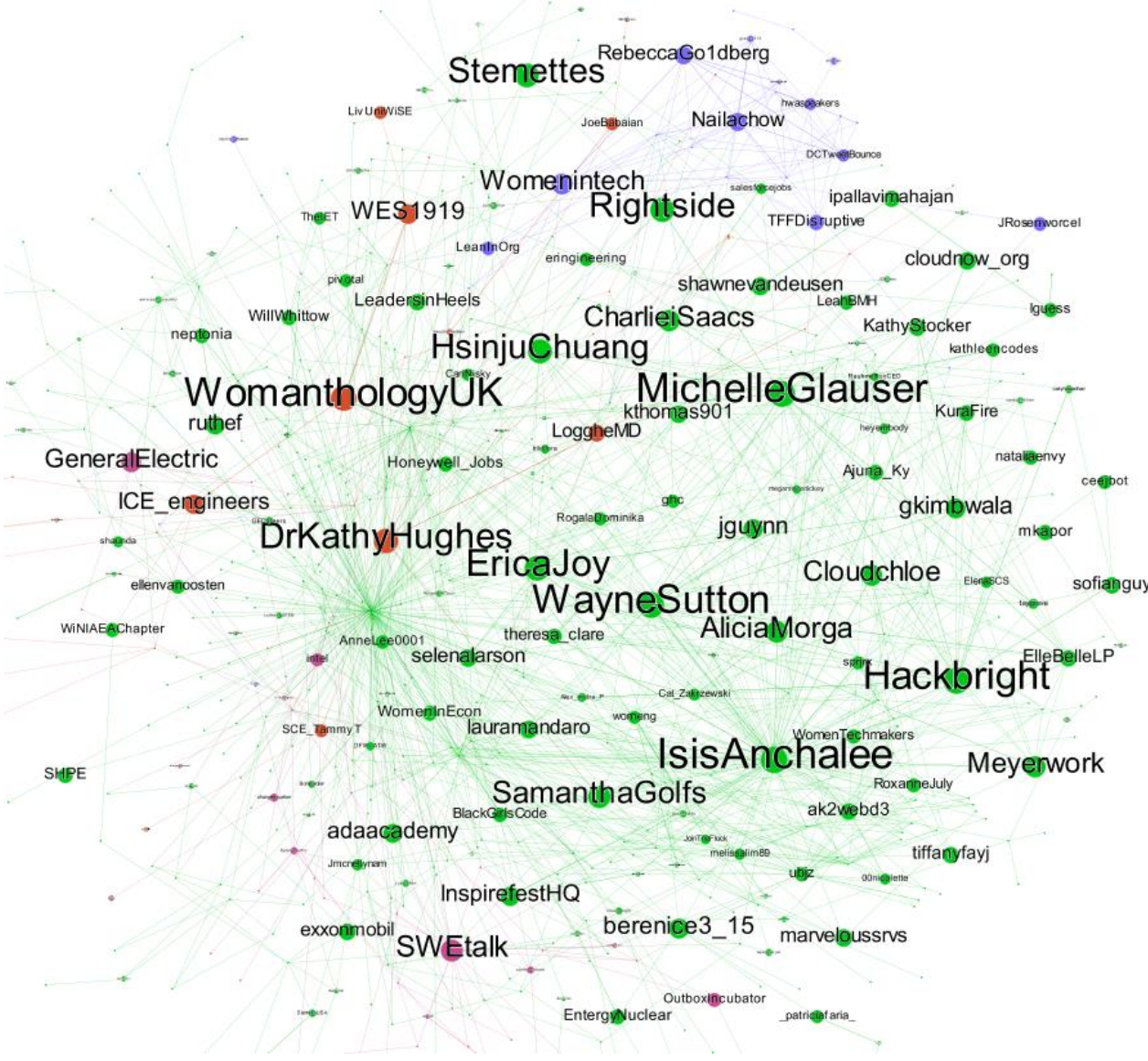

Fig. 4. Network graph of #ILLAE.

In social network analysis two key measures including degree (in-degree and out-degree) and betweenness centrality are used often to assess the key entities within the network. These metrics support in better understanding the involvement of various entities in the conversation. For instance, betweenness centrality measure helps in identifying the most influential entities within a network. The importance of a node within a network is indicated by a high betweenness centrality. On the other hand, users having a high out-degree value indicate the entities responsible for initiating most conversations (retweets in this case), meanwhile users with a high in-degree point out the participants with most of the conversations directed towards them. Node-level metrics of the campaign highlighting the top 10 entities is presented in the Table 5.

Table 5: Entities with highest out-degree, in-degree, and betweenness

| **Highly Gregarious** | | **Highly Referred** | | **Highly Influential** | |
|---|---|---|---|---|---|
| **Twitter handle** | **Out-degree** | **Twitter handle** | **In-degree** | **Twitter handle** | **Betweeness** |
| MichelleGlauser | 71 | EngineerLook | 275 | IsisAnchalee | 26841 |
| IsisAnchalee | 69 | IsisAnchalee | 110 | MichelleGlauser | 10917 |
| CaterpillarInc | 53 | andytoronto | 83 | Hackbright | 9869 |
| GeneralElectric | 37 | liandryaaa | 56 | WayneSutton | 8963 |
| WayneSutton | 29 | SWEtalk | 48 | WomanthologyUK | 7925 |
| Hackbright | 27 | WayneSutton | 42 | EricaJoy | 5634 |
| TechCrunch | 26 | Hackbright | 40 | DrKathyHughes | 4716 |
| austinenergy | 25 | ubjz | 39 | Rightside | 4320 |
| hwaspeakers | 23 | Rosenchild | 38 | HsinjuChuang | 4237 |
| womeng | 23 | KahwaiMuriithi | 30 | Stemettes | 4003 |

Analyzing the entities with high out-degree illustrates the key hubs initiating the conversations within the network. Two females including Michelle Glauser (Campaign Manager) and Isis

Anchalee (Campaign initiator) stand out as the central entities together with large engineering multinationals such as Caterpillar and General Electric. The analysis also reveal two entities including Hackbright Academy and Women in Engineering at Twitter that proactively promote the participation of women in engineering education and workforce. The in-degrees metric analysis indicates that that most of the conversational activity is directed towards the campaign organizers. The official Twitter profile of the #ILookLikeAnEngineer campaign manager and campaign initiator Isis Anchalee are most often referred in the network. Furthermore, the conversational activity was also directed towards two organizations (The Society of Women Engineers and Hackbright Academy) involved in empowerment of women in engineering and technology.

We also calculated betweenness centrality to assess the most influential actors within network. The higher the size of node the more important that actor is in connecting various communities together (see Fig. 4). Likewise, other node-level metrics, Isis Anchalee scored the highest betweenness centrality. The campaign manager Michelle Glauser, and female engineers Erica Joy, Hsin-Ju Chuang also emerged as important players in engaging people in conversation. Finally organization including Hackbright Academy, WomanthologyUK, and Stemettes challenging the stereotypes and promoting women participation in STEM were also considered as the key entities influencing the network.

**Future Work**

In our future work we plan to work with larger datasets and apply techniques such as supervised classification [20], [21] to help us categorize a large set of users automatically. For analyzing interaction patterns of the categorized users, we will create interaction networks [22] by considering nodes as users and edges as the interaction strength—who spreads whose messages, who acknowledges whom, etc. We will use different types of centrality measures and network visualization algorithms, such as Force-Directed layout those available in popular open source tools like Gephi, to identify patterns of clusters, and influential users. This technique will be useful in identifying and understanding interactions of the actors – individuals and organizations – within our dataset that participate in STEM related issues. Topic modeling is a content analysis method to identify hidden topics in the content of discourse and a topic model provides topics as word clusters where the clustered words represent semantically related meaning. For instance, a topic of computer engineering in the discourse might be expressed by use of several engineering related words in different tweet messages—computer, software architecture, scalable computing, cloud backend, system design, development process, etc. We will ultimately use the topical distributions of such word clusters over messages to automatically partition dataset by different topics, and it will help us understand dynamics of participation in the discourse related to diverse STEM issues. Sentiment-emotion mining is a text analysis method to understand subjective behavior of the text author. Among the different variants of the method, a relevant technique to our research is target-specific sentiment estimation, which measures sentiment for a given target in the text. For example, a target can be specific STEM issue of 'women', and therefore, whenever a Twitter message contains words related to the target 'women', this technique provides the valence measure of the message author towards women as positive, negative or neutral with probability. This group of techniques will help us gauge the attitude of participating actors in STEM related discourse towards specific issues. Through this we will be able to develop a deeper picture of what participants feel about STEM.

## Conclusion

We analyzed data from a hashtag activism campaign, #ILookLikeAnEngineer, to identify how the affordances of social media, in our case Twitter, support social movements for STEM diversity. The ability of a diverse range of people to participate in a campaign and to voice their opinions is critical for a broad support to an issue of social good. We find that in the case of this campaign Twitter allowed users to express themselves in different ways but implicitly related, and participate regardless of whether they were an individual or a corporation lending the campaign a strong support and momentum. Our empirical analysis also sheds lights on issues that people consider to be important in order to advance gender diversity within the engineering/technology workplace. Unlike basic observations of the activity of user engagement using counts, a key outcome of such an analysis will be better understanding of patterns of how campaigns and specific issue-based hashtags generate varied types of engagement from the public and where.Overall, our preliminary work provides enough evidence that looking at social media data, Twitter data in our case, provides useful information and when aggregated and thoroughly parsed and analyzed, this information has the potential to provide novel insights related to STEM diversity in education and workforce. We developed the analytical foci for the proposed research after the preliminary work. The preliminary research questions were small in scope but they helped us sharpen our analytical foci to provide a more nuance and useful look at the data.